\documentclass[12pt]{article}
\usepackage{epsfig}

\newcommand{\be}{\begin{equation}}
\newcommand{\ee}{\end{equation}}
\begin{document}

\title{A simple theory for the Raman spike}
\author{J.-G.~Caputo$^{1~2~*)}$ and A.~Maimistov$^3$}
\date{ \  }
\maketitle

{\footnotesize
\hangindent=4mm \noindent
1) Laboratoire de Math\'ematiques, INSA de Rouen,\\
B.P. 8, 76131 Mont-Saint-Aignan cedex, France. \\
\phantom{1)} E-mail: caputo@insa-rouen.fr \\
2) Laboratoire de Physique th\'eorique et modelisation,\\
Universit\'e de Cergy-Pontoise and C.N.R.S.\\
3) Department of Solid State Physics, \\
Moscow Engineering Physics Institute,\\ 
Kashirskoe sh. 31, Moscow, 115409 Russia\\
\phantom{1)} E-mail: maimistov@pico.mephi.ru \\
}

\begin{abstract}

The classical stimulated Raman scattering system
describing resonant interaction between two electromagnetic waves
and a fast relaxing medium wave is studied by introducting a systematic
perturbation approach in powers of the relaxation time. We separate
amplitude and phase effects for these complex fields.
The analysis of the former shows the existence of a stagnation distance
after which monotonous energy transfer begins from one
electromagnetic wave to the other and this quantity is calculated.
Concerning phase effects we give the conditions for the formation of
a Raman spike from an initial fast and large phase jump in one of the
waves. The spike evolution and width estimated from the reduced
model agree with the results from numerical simulations of the 
original system.

\end{abstract}


\section{Introduction}

The investigation of the transient effects in stimulated 
Raman scattering (SRS) has a long history. The initial
interest came from the study of Druhl et al \cite{drulh}
who examined the interaction of a strong laser pump pulse
with a high pressure gas. They found that pulses of long
duration get completely depleted except in some cases 
where a very rapid and short lived restoration occurs. 
Numerical simulations then showed that this
"spike" of pump radiation was connected to a rapid
$\pi$ phase shift or phase flip introduced in the 
initial Stokes wave \cite{wenzel}.

The theory of SRS is based on a very simple system of equations resulting
from the Pluczek model of interaction of electromagnetic waves with
molecular vibrations \cite{yariv,pantel}
This is essentially
the wave equation coupled to an oscillator equation. Using a slowly
varying envelope approximation for the fields and assuming 
the Brillouin resonance conditions
on energy and momentum, one can obtain the system of three wave interaction
between two electromagnetic waves and a medium wave (See, for example 
\cite{cgl95}).
This system is interesting for its mathematical properties and can be 
found in many fields. First, in the limit of zero damping Chu et al
\cite{chu} showed that it can be written as zero-curvature conditions for
two differential operators, so that an inverse scattering 
transform (IST) method \cite{ablowitz} could be used to solve it. However
IST methods are generally used to solve Cauchy problems where one is given
all the field data for all space coordinates at an initial time $t_0$.
The theory of SRS leads to a boundary value problem where the fields
are given for all times at a fixed location $x_0$. As illustrated in
\cite{leon,lm99} the IST method can be developed in this case as
well but one needs to check independently that the potential decays 
sufficiently fast at infinity. This approach was recently validated 
for the model with group velocity dispersion by 
direct comparison with the numerical solution of the problem \cite{bclp00}.

From another point of view the SRS is an example in optics of the very 
important phenomenon of three wave interaction. A second well known
example is the Mandelshctam-Brillouin scattering, where an optical wave
interacts with acoustic phonons. As well as SRS this interaction is
used to amplify directly laser pulses propagating in an optical fiber 
\cite{Agrawal95}, \cite{Boyd}.
The second harmonic generation in non 
collinear beams is another example of three wave interaction. 
Overall this phenomenon is present in much of condensed matter physics, from 
the scattering of spin waves on phonons to the nonlinear optics of 
Bose-Einstein condensates \cite{BEC}. Therefore the results obtained for SRS
could be transposed to these other contexts.

	Here we are interested in the limit 
where the damping of the medium vibrations is large, so that a 
perturbation theory derived from the IST method is not applicable. This
is precisely when one observes the Raman spike, where the pump radiation
after being depleted displays a sharp burst of energy, which lasts a
small fraction of its initial duration. This phenomenon first observed
in gases can be studied via numerical simulations of the full equations
\cite{wenzel,tashkent}. These show that the Raman spike can be seen only for 
large dissipation, that it appears with some delay and that its amplitude 
decays for increasing propagation distances in the medium. Our aim is
to understand how this occurs, more precisely 
\begin{enumerate}
\item understand quantitatively the energy transfer from one component of 
the field (the pump) to the other (Stokes wave), in particular estimate 
the distance necessary for a given transfer.
\item explain the time delay of the Raman spike (stagnation effect)
\item explain the position, the width and the amplitude of the 
Raman spike as a function of the initial phase flip in the Stokes wave
\item understand the evolution of the fronts of the pump pulse
\end{enumerate}

We consider a simplified situation where the pump and Stokes pulse are 
not gaussian but almost rectangular. In this case we will see that the
effects of phase and amplitude can be separated so that the magnitude of
the wave intensities can be assumed to be independent of time in the
spike region, located in the middle of the pulses. The medium variable can be
expressed as an integral of the product of the amplitudes of the interacting
waves. This can be expanded using as a small parameter $1 /\Gamma$ where
$\Gamma$ is the damping of the medium variable. If we limit ourselves to the 
leading term in this series, we obtain the well-known equations describing
SRS for continuous waves and within this approximation we find the
Raman transfer coefficient. At the next order we obtain a system of equations
where the evolution of the pump amplitude is independent from the one of
the phase. Though it cannot take into account the phase flip effects 
causing the Raman spike, it should be adequate for the transient 
phenomena occurring on the fronts
of the interacting pulses. To get coupling between 
the phase and amplitudes of the pulses we need to consider
the third order term in the expansion of the medium variable.  
In this approximation the dispersion of the response is taken into account and
this causes phase variations to influence the amplitudes.
Thus, we have the simplest approximate theory to describe the Raman 
spike phenomenon. We will show that nevertheless our analytical 
description of the Raman spike formation reproduces the stagnation effect 
and yields an approximate formula for the spike width. The comparison 
of our analytical results with 
the results of the numerical simulation shows qualitative and 
in some cases quantitative agreement.

The paper is organized as follows. After introducing the original
model in section 2, we describe the approximation used and obtain
the zero order (standard Raman model) in section 3, the front effects
corresponding to the first order are described in section 4. Dispersion
effects corresponding to the third order are given in section 5 and
analyzed in section 6. We present numerical results confirming this
approach in section 7 and conclude in section 8.

\section{The original model }

The system of equations describing the non-stationary stimulated Raman
scattering can be written as \cite{newell}
\begin{equation}
\frac{\partial a}{\partial x}=qb,\quad \frac{\partial b}{\partial x}%
=-q^{\ast }a,  \label{eq11}
\end{equation}

\begin{equation}
\frac{\partial q}{\partial t}+\Gamma q=-gab^{\ast },  \label{eq12}
\end{equation}

\noindent where $a,b$ are the amplitudes of the electromagnetic field and
$q$ is the amplitude of the medium. 
$\Gamma $\ is the damping coefficient and $g$ is the amplification 
coefficient. This system of equations may be presented in a dimensionless 
form by rescaling $a,b$ and taking as unit of time the
dissipation time $1/\Gamma$,
$ a = \tilde{a}/g^{1/2}$, $b = \tilde{b}/g^{1/2}$ and  $t =\tilde{t}/\Gamma$.

\noindent One gets, omitting the tildes
\begin{equation} 
\frac{\partial a}{\partial x}=qb,\quad \frac{\partial b}{\partial x}%
=-q^{\ast }a,  \label{ab}
\end{equation}
\begin{equation}
\frac{\partial q}{\partial t}+q=-\varepsilon ab^{\ast }  \label{qq}
\end{equation}
where $\varepsilon =1/\Gamma$ is a small parameter because we consider the
case of a large dissipation.

The initial conditions for the system are
$$x=0, a= a_0(t), b=a_0(t) \rho \exp{i\phi(t)}, $$
$$t=0, q= 0$$
where the last equation indicates that the medium is 
initially in the ground state.
The quantity $\rho = e^{-\gamma}$ is the initial amplitude difference
between $a$ and $b$, in most cases we will assume $ \rho << 1$ and 
$|a|_{x=0}$ to be of order 1 so that $|a|_{x=0} >>  |b|_{x=0}$.
The term $\phi(t)$ is a time dependent initial phase.
In the following we will assume that $a_0, \gamma$ and $\phi$ are real.

Equations (\ref{ab}) and (\ref{qq}) are such that $D^2(t)\equiv |a|^2 +|b|^2$ is
independent of $x$ and therefore $D^2(t)=|a|^2 +|b|^2=a_0^2(t)  (1+\rho^2)$.
There is no symmetry in exchanging $a$ and $b$. In fact
the evolution of the field variable $q$ can be written
$$ q_{xt} + q_x =\epsilon q (|a|^2 -|b|^2).$$
Even if $|a|^2 -|b|^2|_{x=0} <0$ there will be no flow of energy from
$b$ to $a$. Note also that an initial homogeneous phase in $b$ plays
no role in the dynamics.

We discretized (\ref{eq11},\ref{eq12}) in both $x$ and $t$ using Heun's
order 2 Runge-Kutta method and advance via the following algorithm 
\begin{enumerate}
\item given $a(x,t)$ and $b(x,t)$ for a given $x$ compute $q(x,t) $
by integrating (\ref{eq12}) in $t$ for the initial datum $q(x,t=0)=0$.
\item advance to $a(x+dx,t)$ and $b(x+dx,t)$ for all $t$ by 
integrating (\ref{eq11}) in $x$.
\item go to step 1 with $x=x+dx$.
\end{enumerate}
The scheme is started at step 1 for $x=0$ and the quality of the
computation is monitored by evaluating the relative error in $|a|^2 +|b|^2$.
In all the runs presented it remained smaller than $10^{-5}$.

\section{A systematic approximation: the zero order}

The equation describing the evolution of the variable $q$ contains the product
of the amplitudes of the interacting fields. This suggests 
to use the variables $\sigma =ab^{\ast }$ and $ n=a^{\ast }a-b^{\ast }b$, 
so that the principal equations take the form
\begin{equation}
\frac{\partial \sigma }{\partial x}=-qn,\quad \frac{\partial n}{\partial x}%
=2(q\sigma ^{\ast }+q^{\ast }\sigma ),  \label{eq21}
\end{equation}
\begin{equation}
\frac{\partial q}{\partial t}+q=-\varepsilon \sigma  \label{eq22}
\end{equation}
The formal integration in (\ref{eq22}) leads to
\begin{equation}
q=-\varepsilon \left( \sigma -\frac{\partial \sigma }{\partial t}+\frac{%
\partial ^{2}\sigma }{\partial t^{2}}-\frac{\partial ^{3}\sigma }{\partial
t^{3}}+...\right),  \label{eq23}
\end{equation}
\noindent which written as an infinite expansion is an
exact solution of (\ref{qq}) that can be obtained by integration 
by parts. 

\noindent To see this consider the formal solution of (\ref{qq}) 
$$q(t)=-\epsilon \int_{-\infty }^{t}
\exp [-(t-t^{\prime })]\sigma (t^{\prime })dt^{\prime } . $$
Integrating by parts, this expression can be rewritten as 
$$q(t)=-\epsilon \left( \sigma (t)-\int_{-\infty }^{t}
\exp [-(t-t^{\prime})]
{\partial \sigma (t^{\prime }) \over \partial t^{\prime}}dt^{\prime }
\right) , $$
or, after a second step, 
$$q(t)=-\epsilon \left( \sigma (t)-{\partial \sigma (t)\over \partial t}
+\int_{-\infty}^t \exp [-(t-t^{\prime })]
{\partial^{2}\sigma (t^{\prime }) \over \partial t^{\prime 2}}dt^{\prime }
\right) ,  $$
and so on. 

\noindent For $t=\Gamma $  we have 
$\partial \sigma (t)/\partial t=O(\sigma /\Gamma )$
Thus this expansion becomes a better approximation to the solution
when the dissipation parameter gets large.

In the zero order approximation, the medium variable follows
instantaneously the field variables
$$q = - \epsilon \sigma .$$
This is a well known approximation \cite{Agrawal95} where the intensities
$|a|,|b|$ can be calculated exactly. 
We get
\begin{equation}\label{a20}
|a|^2 (x,t)=  D^2(t) {1   \over 1+\rho^2 e^{2 \epsilon D^2 x} },
\end{equation}
\begin{equation}\label{b20}
|b|^2 (x,t)=  D^2(t) {\rho^2 e^{2 \epsilon D^2 x}  \over 
1+\rho^2 e^{2 \epsilon D^2 x} } ~~~.
\end{equation}
Here we introduce $\rho = e^{-\gamma}$ so that the
key quantity in (\ref{a20},\ref{b20}) 
$$\rho^2 e^{2 \epsilon D^2 x} = e^{2(\epsilon D^2 x -\gamma)}.$$
We immediately see that the distance $x_c$ 
past which a significant transfer of energy occurs is 
\be\label{stag}
x_c= {\gamma \over \epsilon D^2 } , \ee
which we will refer to as the stagnation distance. 
We will find this quantity throughout this work.

To measure energy transfer between modes $a$ and $b$, it is convenient 
to introduce the Raman transfer coefficient
\be  \label{transfer}
R(x) = 1 - {\int_{-\infty}^{+\infty} |a|^2 dt 
\over \int_{-\infty}^{+\infty} |a_0|^2 dt} ,
\ee
such that $0 < R(x) <1$.
Initially $a= a_0$ so that $R(x=0)=0$. As $x$ increases $a$
decreases so that $R$ increases towards 1.
Assuming $a_0$ to be a rectangular initial pulse 
$|a_0|^2 (t) = H(t-t_1) -H(t-t_2)$ where H is the usual
Heaviside function, we get $\int_{-\infty}^{+\infty} |a_0|^2 dt= t_2 - t_1$
and 
$$\int_{-\infty}^{+\infty} |a|^2 dt= (t_2 - t_1)
{1+\rho^2 \over \rho^2 e^{2 (1 + \rho^2)\epsilon x} +1}$$
so that
\be  \label{trans}
R(x) = \rho^2 {e^{2 (1 + \rho^2)\epsilon x} -1 \over 
\rho^2 e^{2 (1 + \rho^2)\epsilon x} +1}
\ee

\begin{figure}[t]
\centerline{\psfig{figure=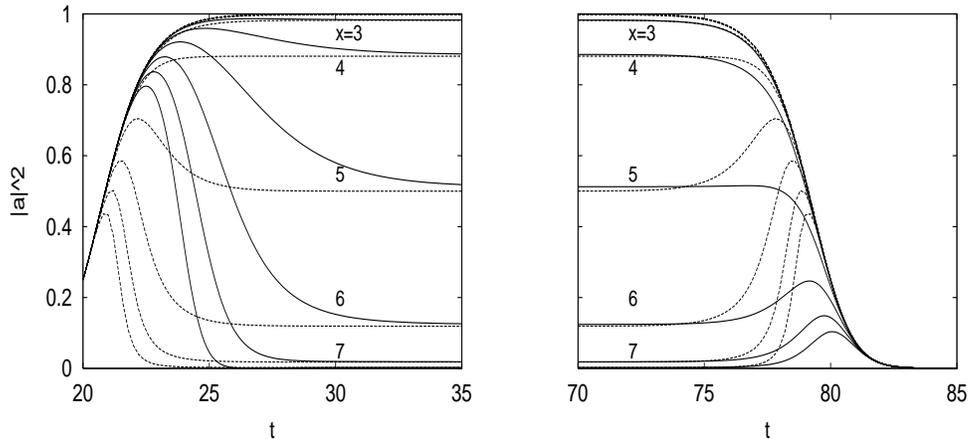,height=15 cm,width=8 cm,angle=-90}}
\caption{Plot of $|a(x,t)|^2$ as a function of $t$ for different
values of $x$ for the numerical solution (full line) and the 
zero order approximation (\ref{a20}). The parameter 
$\gamma =5$.} 
\label{f1} \end{figure}

In Fig. \ref{f1} we plot $|a(x,t)|^2$ as a function of time $t$
for $x=3,4,5,6$ and $7$ for both the solution of the partial
differential equation system (\ref{ab},\ref{qq}) shown in
full line and the 
zero order approximation (\ref{a20}) shown in dashed line. 
The initial pulse is given by
\be\label{a0}
a_0(t) = 0.5 (\tanh((t-t_i)/w_t)-\tanh((t-t_e)/w_t) )  ,\ee
\noindent where $t_i=20$ and $t_e=80$ are respectively the front and back ends
of the pulse and the front width $w_t=2$. The 
left (resp. right) panel of Fig. \ref{f1} corresponds to $20<t<35$ 
(resp. $70<t<85$). 
The zero-order approximation predicts correctly the field depletion
in the center of the pulse but fails for the values of $t$ corresponding
to the incoming or outgoing fronts.

\begin{figure}[h]
\centerline{\psfig{figure=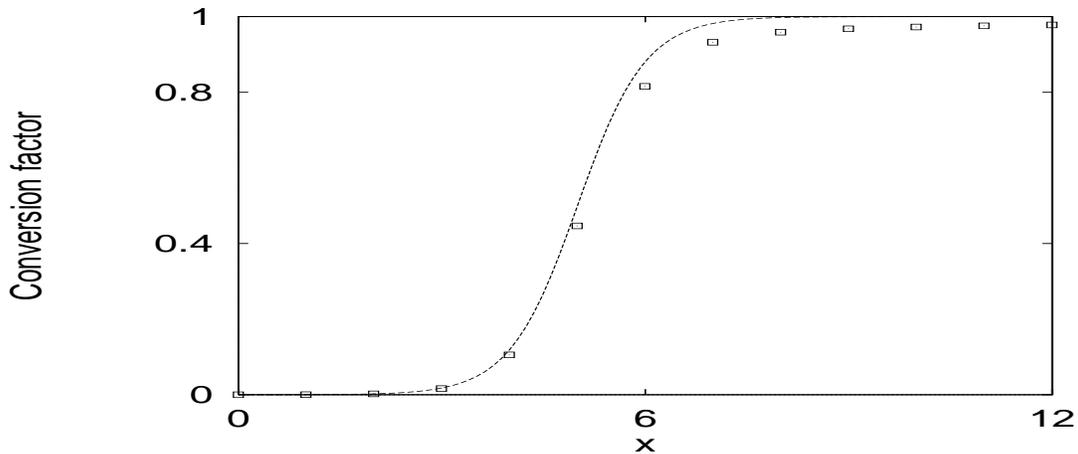,height=15cm,width=6 cm,angle=-90}}
\caption{Computation of the Raman transfer coefficient vs. the position
$x$ for the numerical
solution of (\ref{ab},\ref{qq}), shown as points and the analytical
expression (\ref{trans}) assuming a Heaviside distribution $a_0(t)$,
shown in full line. The parameter $\gamma=5$.}
\label{f2} \end{figure}

Despite this disagreement on the fronts, the estimation (\ref{trans}) of the 
Raman transfer coefficient shown in Fig. \ref{f2} matches fairly well
the one obtained from the full numerical solution. Since the pulse
is long compared with the typical time $1/\Gamma$ (=1 here) 
the errors made at the front ends are compensated and the
approximation is good. This would not be the case for a short
pulse.

It seems clear that the transient effects occurring near an incoming
or outgoing pulse cannot be captured by the zero order approximation
and that higher order terms are needed. In the next section we
include these higher order terms limiting ourselves to the
first order time derivative.

\section{The first order: front effects}

We substitute the corresponding approximation for $q$\ in 
the initial system of equation to obtain
\begin{equation}
\frac{\partial \sigma }{\partial x}=n\left( \sigma -\frac{\partial \sigma 
}{%
\partial t}\right) ,  \label{eq24}
\end{equation}
\begin{equation}
\frac{\partial n}{\partial x}=-2\left( 2\left| \sigma \right| ^{2}-\frac{%
\partial \left| \sigma \right| ^{2}}{\partial t}\right) .  \label{eq25}
\end{equation}
We introduce the real variable $p$ according to the definition 
$p=2\sigma \exp (-i\phi )$
so that we now only have real variables which satisfy the system
of equations 
\begin{equation}
\frac{\partial p}{\partial x}+n\frac{\partial p}{\partial t}=np,
\label{eq26}
\end{equation}
\begin{equation}
\frac{\partial n}{\partial x}-\frac{1}{2}\frac{\partial p^{2}}{\partial t}%
=-p^{2},  \label{eq27}
\end{equation}
\begin{equation}
\frac{\partial \phi }{\partial x}+n\frac{\partial \phi }{\partial t}=0.
\label{eq28}
\end{equation}
If we remember that $\sigma =ab^{\ast }$, one finds that $\phi =\arg a-\arg 
b$, and for $x=0$, $\phi $\ is the function which determines the
phase flip process. The system of equations given above admits the 
conserved quantity with respect to $x:$%
\begin{equation}\label{p2pn2}
p^{2}+n^{2}=D^{4}(t)=\left[ a_{0}^{2}(t)+b_{0}^{2}(t)\right] ^{2}
\label{eq29}
\end{equation}
This relation allows us consider only two equations, one for the difference of
the squares of the interacting waves and another for the
phase difference $\phi $.
\begin{equation}\label{fren}
\frac{\partial n}{\partial x}+n\frac{\partial n}{\partial t}=2D^{3}\frac{%
\partial D}{\partial t}+(n^{2}-D^{4}),
\end{equation}
\begin{equation}\label{frep}
\frac{\partial \phi }{\partial x}+n\frac{\partial \phi }{\partial t}=0.
\end{equation}

\section{Analysis of the characteristic equations for an incoming front}

We now proceed to study the effect of an incoming front using 
equations (\ref{fren},\ref{frep}). We first simplify the right
hand side of the equation for $n$ by writing it $n^2 - f(t)$ where
$f(t)\equiv D^3(D - 2 {\partial D \over \partial t})$.
Then we write the equations in characteristic form
\be \label{frcar}
{dn \over dx}=n^2- f(t)      ~~~~~~~~; ~~~~~~~~ {d t \over dx}=n  ~~.
\ee
The equation for $n$ can be integrated using separation of variables
and we obtain
\be \label{frn} n = - \sqrt{f(t)} \tanh [\sqrt{f(t)} (x-x_0) ] , \ee
and 
\be \label{frt}
t=t_{0}+\int\limits_{0}^{x}n(\xi )d\xi = t_{0} 
-\log[{\cosh(\sqrt{f(t)} (x-x_0)) \over \cosh(\sqrt{f(t)} x_0)}] ,  
\ee
\noindent where the square root can be imaginary so that the
$i \tanh$ becomes a $\tan$.
The integration constant $x_0$ can be computed from the 
initial condition and the relation $ n_0/D^2 = \tanh(\gamma)$
obtained from the definition of $D$.
We get
$$x_0 = {1 \over f}{\rm atanh}[{D^2 \over f}\tanh(\gamma)] ~~.$$
The expressions (\ref{frn},\ref{frt}) give $n$ and $t$ in implicit form
and therefore need to be solved numerically.

\begin{figure}[t]
\centerline{\psfig{figure=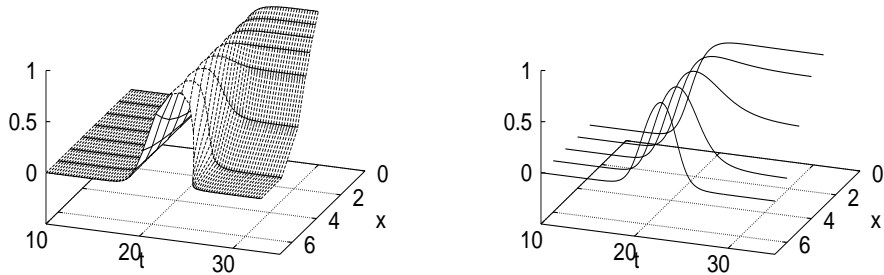,height=15cm,width=6 cm,angle=-90}}
\caption{3D plot of the incoming front of Fig. \ref{f1}. The left
panel shows the solution obtained from the characteristics of the
approximate model (\ref{frcar}) while the right panel gives the full
numerical solution of (\ref{ab},\ref{qq}).}
\label{f3} \end{figure}

We have integrated numerically  the characteristic ordinary differential
equations (\ref{frcar}) in $x$ and obtained the solution in the form
of characteristics $t(t_0,x),n(t_0,x)$. In Fig. \ref{f3} this
estimation is 
compared to the full numerical solution of (\ref{ab},\ref{qq}) for the incoming
front of Fig. \ref{f1}. The results are shown in
the form of 3D plots $|a(x,t)|^2$, on the left panel is the solution
given by the characteristics while the full numerical solution is
given in the right panel. For large $x$ the plots on the left panel 
are curved because the wave breaks i.e. it becomes multivalued. The
approximate model (\ref{frcar}) is therefore not sufficient to
describe an incoming front. Second derivative terms should be included
to yield a dynamics similar to the one of Burgers equation.

\section{Dispersion effects: second order approximation}

Lets now consider the terms of (\ref{eq23}) up to the second order with respect
to the time derivative and rescale the independent variable $x$ into
$ x^{\prime }=\varepsilon x$. The substitution
of expression
\begin{equation}
q=-\varepsilon \left( \sigma -\frac{\partial \sigma }{\partial t}+\frac{%
\partial ^{2}\sigma }{\partial t^{2}}\right)  \label{eq31}
\end{equation}
into the equations (\ref{eq21}) results in the following system of equations:
\begin{equation}
\frac{\partial \sigma }{\partial x}=n\left( \sigma -\frac{\partial \sigma 
}{%
\partial t}+\frac{\partial ^{2}\sigma }{\partial t^{2}}\right) ,
\end{equation}
\begin{equation}
\frac{\partial n}{\partial x}=-2\left( 2\left| \sigma \right| ^{2}-\frac{%
\partial \left| \sigma \right| ^{2}}{\partial t}+\frac{\partial ^{2}\left|
\sigma \right| ^{2}}{\partial t^{2}}-2\left| \frac{\partial \sigma }{%
\partial t}\right| ^{2}\right) ,  \label{eq32}
\end{equation}
where the primes have been dropped for simplicity.
Both these equations and the original system of equations (\ref{eq21}) 
and (\ref{eq22}) have the first integral with respect to $x$
$n^{2}+4\sigma ^{\ast }\sigma =const(t)$.

\noindent Once again we introduce the real variable $p$ according to the
definition
$2\sigma =p\exp (i\phi )$. Using the system of equation (\ref{eq31}) and 
(\ref{eq32}) one gets the following system of equations for the real variables
$p,n$ and$\phi$
\begin{equation}
\frac{\partial p}{\partial x}+n\frac{\partial p}{\partial t}-n\frac{\partial
^{2}p}{\partial t^{2}}=np[1-\left( \frac{\partial \phi }{\partial t}\right)
^{2}],  \label{eq33}
\end{equation}
\begin{equation}
\frac{\partial n}{\partial x}-p\frac{\partial p}{\partial t}=
-p^2 [1-   \left( \frac{\partial \phi }{\partial t }\right)^{2}]  
- p \frac{\partial^{2}p}{\partial t^{2}}
,  \label{eq34}
\end{equation}
\begin{equation}
\frac{\partial \phi }{\partial x}+n\frac{\partial \phi }{\partial 
t}-n\frac{%
\partial ^{2}\phi }{\partial t^{2}}=2\frac{\partial \phi }{\partial t}\left(
\frac{\partial \log p}{\partial t}\right)n ,  \label{eq35}
\end{equation}
where again the conservation law (\ref{p2pn2}) holds.

Since we want to take into account only effects resulting from the phase flip
process, we will neglect the terms ${\partial \log(p)  \over \partial t} $ and 
${\partial^2 p  \over \partial t^2} $. 
Furthermore, we will assume that $D$ does not vary with time, i.e. we place
ourselves in the central part of the rectangular-like initial pulse. 
Using these approximations the system of equations 
(\ref{eq33}) -- (\ref{eq35}) takes the form: 
\begin{equation}
\frac{\partial n}{\partial x}+n\frac{\partial n}{\partial t}%
=(n^{2}-D^{4})\left[ 1-( \frac{\partial \phi }{\partial t})^{2}
\right ],  \label{eq37}
\end{equation}
\begin{equation}
\frac{\partial \phi }{\partial x}+n\frac{\partial \phi }{\partial t}=0.
\label{eq38}
\end{equation}
These equations can be written in the characteristic form \cite{w74}:
\begin{equation}
\frac{dt}{dx}=n,\quad \frac{d\phi }{dx}=0,  \label{eq39}
\end{equation}
\begin{equation}
\frac{dn}{dx}=(n^{2}-D^{4})\left[ 1-\left( \frac{\partial
\phi }{\partial t}\right) ^{2}\right ]  \label{eq310}
\end{equation}
From equations (\ref{eq39}) it follows that $\phi =\phi (t_{0})$\ and the
expression to find characteristics
\begin{equation}
t=t_{0}+\int\limits_{0}^{x}n(\xi )d\xi ,  \label{eq311}
\end{equation}
where $\xi $ is the variable of the characteristic. From (\ref{eq310}) 
we can compute $n(\xi )$\ as 
\begin{equation}
n(x,t_{0})=-D^{2}\tanh \left\{ D^{2}f(t_{0})\left[ x-x_{0}(t_{0})\right]
\right\} ,  \label{eq312}
\end{equation}
where $x_{0}(t_{0})$ is a constant of integration and the function 
$f(t_{0})$ is defined by 
\[
f(t_{0})=1-\left( \frac{\partial \phi }{\partial t}\right) ^{2}
\]
where the derivative with respect to $t$ is calculated at $t=t_{0}$, and 
as $\phi =\phi (t_{0})=\phi (x=0,t_{0})$, 
its value can be calculated at $x=0$. Thus the function $f(t_{0})$ 
reflects the phase flip on the initial pulse $b$ of the interacting
waves. The constant of integration $x_{0}(t_{0})$\ is defined by the
initial condition
\begin{equation}
n(x=0,t_{0})=n_{0}=D^{2}\tanh \left\{ D^{2}f(t_{0})x_{0}(t_{0})\right\} .
\label{eq313}
\end{equation}
If we use equation (\ref{eq313}) and the definitions of $n$\ and $D$ we obtain:
\[
n_{0}=a_{0}^{2}\left( 1-\exp (-2\gamma )\right) ,\quad D^{2}=a_{0}^{2}\left(
1+\exp (-2\gamma )\right) ,
\]
then the constant of integration $x_{0}(t_{0})$ can be rewritten as $%
x_{0}(t_{0})=\gamma \left[ D^{2}f(t_{0})\right] ^{-1}$. Substituting this
expression into equation (\ref{eq312}) leads to
\begin{equation}
n(x,t_{0})=-D^{2}\tanh \left\{ D^{2}f(t_{0})x-\gamma \right\} .
\label{eq314}
\end{equation}

So equations (\ref{eq311}) and (\ref{eq314}) allow
us to write the expression defining the characteristics
\begin{equation}
t=t_{0}-\frac{1}{f(t_{0})}\log \left\{ \frac{\cosh \left[ \gamma
-D^{2}f(t_{0})x\right] }{\cosh \gamma }\right\} .  \label{eq315}
\end{equation}
From this equation one can extract $t_{0}$ as a function of 
$t$ and $x$ and substitute it into
(\ref{eq314}) to yield the solution of the problem
under consideration:
\begin{equation}
n(x,t)=-D^{2}\tanh \left\{ D^{2}f(t_{0}(t,x))x-\gamma \right\} ,
\label{eq316}
\end{equation}
\begin{equation}
a^{2}(x,t)=0.5D^{2}\left[ 1-\tanh \left\{ D^{2}f(t_{0}(t,x))x-\gamma
\right\} \right] .  \label{eq317}
\end{equation}
There is an intrinsic difficulty in this procedure due to the fact that
the characteristics can cross so that there might not be a unique $t_{0}$
for a given $t$ and $x$. In particular we will see that around the spike
region characteristics will cross. Nevertheless it is possible to understand
what is going on by using asympotics in the regions away from where the
characteristics cross. We do this in the next section. 

\begin{figure}[t]
\centerline{\psfig{figure=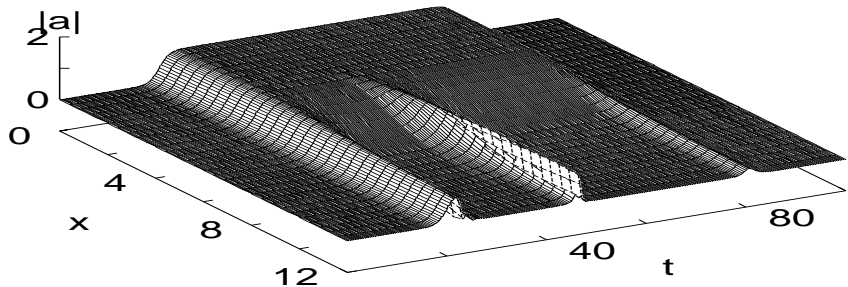,height=15 cm,width=6 cm,angle=-90}}
\caption{3D plot of the pump amplitude $|a| (x,t)$ obtained
by direct integration of the original equations (\ref{ab},\ref{qq}).
The initial phase flip is for $t=t_f=40$ and the
parameters are $\gamma=5$ and $\delta=0.011$.}
\label{f4} \end{figure}

\section{Analysis of the characteristic equations for the Raman spike} 

As shown by expressions (\ref{eq315}) and (\ref{eq317}) the principal
problem is to determine the auxiliary function $t_{0}(t,x)$, from
the equation of characteristics (\ref{eq315}). 
To do it we will use the numerical solution of this equation. 
However, it is possible to obtain some information for small or
large $x$ using asymptotic expansions of the function $t_{0}(t,x)$. 
We write
\be\label{tt0} t = t_0 + {1\over f}\log[\cosh(D^{2}~f~x)] 
+{1\over f}\log [1-\tanh(D^{2}~f~x)~\tanh(\gamma)]\ee
For $x$ small the above expression can be expanded to yield
at first order
\be\label{txsmall}
t \approx t_0 + D^{2}~x \tanh(\gamma)= t_0 + n_0 x~.~\ee
This expression can be immediately deduced from the characteristic
equation giving $dt /dx$ (\ref{eq311}).
It is correct even in the spike region as long as
$D^{2}~f~x<<1$.
For larger values of $x$ it is necessary to distinguish the
region where $f<<0$ leading to a spike from the rest of the
domain where $f\approx 1$.

Let us consider this case first. Then one can approximate the
expression (\ref{tt0}) for large $x\approx 2 {\gamma \over D^{2}}$ by
\be\label{txlarge}
t = t_0 +2 \gamma \tanh(\gamma)  - D^{2}~x \tanh(\gamma)~.~\ee
Notice how the slope of the $t(x)$ curve, i.e. the 
direction of the characteristic changes as one crosses the stagnation
distance $x_c= {\gamma \over D^{2}}$.

The situation in the spike region can be analyzed using the expression
for $n$ (\ref{eq316}). Inside the spike region the argument of the
tanh in $n$, $D^{2}~f~x -\gamma <0$ even for large values of $x$
so that there is no depletion. In this region the characteristics follow
$t = t_0 + D^{2}~x \approx t_0 + n_0 x~$ which is very close to (\ref{txsmall})
for large $\gamma$. On the sides of the spike $0<f<1$ so that the argument of
the tanh can go through 0 but for a larger stagnation distance than
$x_c$. To summarize, outside the spike region we expect the characteristic 
curves to bend smoothly from the behavior (\ref{txsmall}) 
to (\ref{txlarge}) as $x$ increases while inside the spike region they will
follow (\ref{txsmall}). There is also an intermediate region around the
spike where the characteristic curves will shift to (\ref{txlarge}) but
for $x>x_c$. These sets of characteristic curves will cross as we will
show in the next section, leading to shock formation.

To understand things quantitatively let us now give a precise form of the
phase flip function. We chose 
\[
\phi (x=0,t)=-\frac{a_f }{2}\left[ 1+\tanh \left( 
\frac{t-t_{f}}{%
\delta }\right) \right] .
\]
\noindent where $t_f$ is the location of the phase jump and $a_f$
its amplitude. When $a_f=\pi$ we speak of phase flip.
With this choice the function $f(t_{0})$ takes the form
\begin{equation}
f(t_0)=1-\frac{a_f^2}{4\delta^2} {\rm sech}^{4} \left( 
\frac{t_0-t_f}{\delta }\right) .  \label{eq44}
\end{equation}
The region where the function $f(t_{0})$ has a minimum will be named
\emph{Raman spike} (RS) \emph{region}. If the parameter $\delta $ is 
chosen to be small, the RS-region is very narrow. Outside this region 
$f(t_{0})$ is equal to one. Hence we can write
\begin{equation}
a^{2}(x,t)=0.5D^{2}\left[ 1-\tanh (D^{2}x-\gamma)  \right]
\label{eq45}
\end{equation}
It shows that up to $x\approx (\gamma -1)/D^{2}$ the value of 
the intensity of the pump wave is $a^{2}(x,t)=D^{2}$ and that for 
$x=x_{c}\approx \gamma /D^{2}$ this value decreases down to 
$0.5D^{2}$. Only inside the RS-region the intensity of the
pump wave does not vary. This results in the formation of a 
Raman spike. 
Fig. \ref{f4} shows a 3D plot of the pump field $|a(x,t)|$ 
solution of the original equations (\ref{ab},\ref{qq}) where we chose
the Heaviside-like initial pump and Stokes distribution (\ref{a0})
In this case the middle of the pulse is such that $D = 1$. We took
$\gamma=5$ and clearly see in Fig. 2 that the pump field starts getting
depleted only past $x_c= 5$. Thus our estimation of the stagnation distance
is a good one.

Let $\Delta _{RS}$ be the width of the RS-pulse given by 
half of the amplitude, i.e. 
$ \Delta _{RS}=2(t_{1}-t_{m})$, where the instant $t_{1}$ is
defined by the condition $a^{2}(x,t_{1})=0.5D^{2}$ and $t_m$ is the
position of the minimum of the function $f(t_0)$. 
Using (\ref{eq317}) and (\ref{eq44}) we can obtain the following equation
\begin{equation}
\frac{a_f^{2}}{4\delta ^{2}}{\rm sech}^4\left( \frac{%
t_{0}(t_{1},x)-t_{f}}{\delta }\right) =1-\frac{\gamma 
}{D^{2}x}=1-\frac{x_{c}%
}{x}.  \label{eq46}
\end{equation}
Note how \emph{the stagnation distance} $x_c$ appears again. 
Furthermore if $x\leq x_{c}$ the right hand side is negative
so that no value of $t_0$ can be found and there is not 
Raman spike formation. Hence we obtain the stunted 
growth (stagnation) of Raman transformation once again. 
From the definition of the stagnation distance $x_{c}$ 
one can see that in dimensional variables $x_{c}$ 
is proportional to the damping coefficient $ \Gamma $ and 
to the initial depletion coefficient $\gamma $.

Let us now proceed to further analyze (\ref{eq46}) in order
to estimate the width of the Raman spike. For that denote
\[
\epsilon ^{4}=\frac{4\delta ^{2}}{a_f^{2}}
\left( 1-\frac{x_{c}}{x}\right)  ,
\]
so that equation (\ref{eq46}) can be written as 
\begin{equation}
\cosh \left( \frac{t_{0}(t_{1},x)-t_{f}}{\delta }\right) =
\epsilon ^{-1} .
\label{eq47}
\end{equation}
Since the parameter $\epsilon $ is very small because of the
smallness of $\delta$, the argument of the $\cosh$ function 
in (\ref{eq47}) is large so that the function can be approximated
by an exponential. We obtain
\[
\exp \left( \frac{t_{0}(t_{1},x)-t_{f}}{\delta }\right) =
\frac{2}{\epsilon } ,
\]
so that $t_{0}(t_{1},x)=t_{f}+\delta \log (2/\epsilon )$. If 
the function $t_{0}(t,x)$ can be approximated by 
$t_{0}\approx t-n_{0}x.$, then $t_{1}=t_{f}+n_{0}x+\delta 
\log (2/\epsilon )$. But in the framework of the
approximation under consideration $t_{m}=t_{f}+n_{0}x$, hence $\Delta
_{RS}\approx 2\delta \log (2/\epsilon )$. From this expression and the
definition of the parameter $\epsilon $\ one gets
\begin{equation}\label{wrs}
\Delta _{RS}\approx -\frac{\delta }{2}\log \left[ 
\frac{\delta ^{2}}{4 a_f^{2}}\left( 1-\frac{x_{c}}{x}\right) \right ] 
\label{eq48}
\end{equation}
where it was assumed that $\delta ^{2}(x-x_{c})<<x_{c}$, 
i.e.,. the distance away from $x_c$ is small large 
enough. It is interesting that for 
$ x\longrightarrow \infty $\ the width of RS-pulse approaches
the limiting value
\[
\Delta _{RS}\approx \delta \log \left( 2 a_f /\delta \right) .
\]
For $\delta =0.5$ this limit is equal to $1.27$. When 
the distance $x$\ approaches the stagnation distance 
$x_{c}$, the width of the RS-pulse increases. Hence the Raman
spike forms only if $x>x_{c}$, a feature that we can clearly see
in Fig. \ref{f4}. 

\begin{figure}[t]
\centerline{\psfig{figure=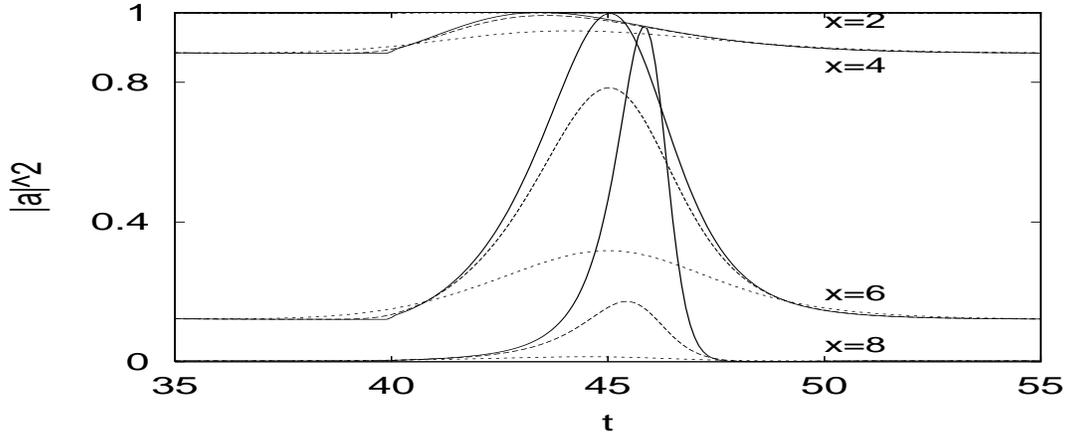,height=15 cm,width=6 cm,angle=-90}}
\caption{Influence of flip time on the formation of the Raman spike.
$|a|^2 (x,t)$ for $x =2,4,6$ and $8$  as a function of $t$ for
three different flip times $\delta = 0.1$ (full line), $1$ (long dash)
and $3$ (short dash). The amplitude of the flip is $\pi$ and the other
parameters are as in Fig. \ref{f4}.}
\label{f5} \end{figure}

\section{Numerical results and discussion}

\begin{figure}[t]
\centerline{\psfig{figure=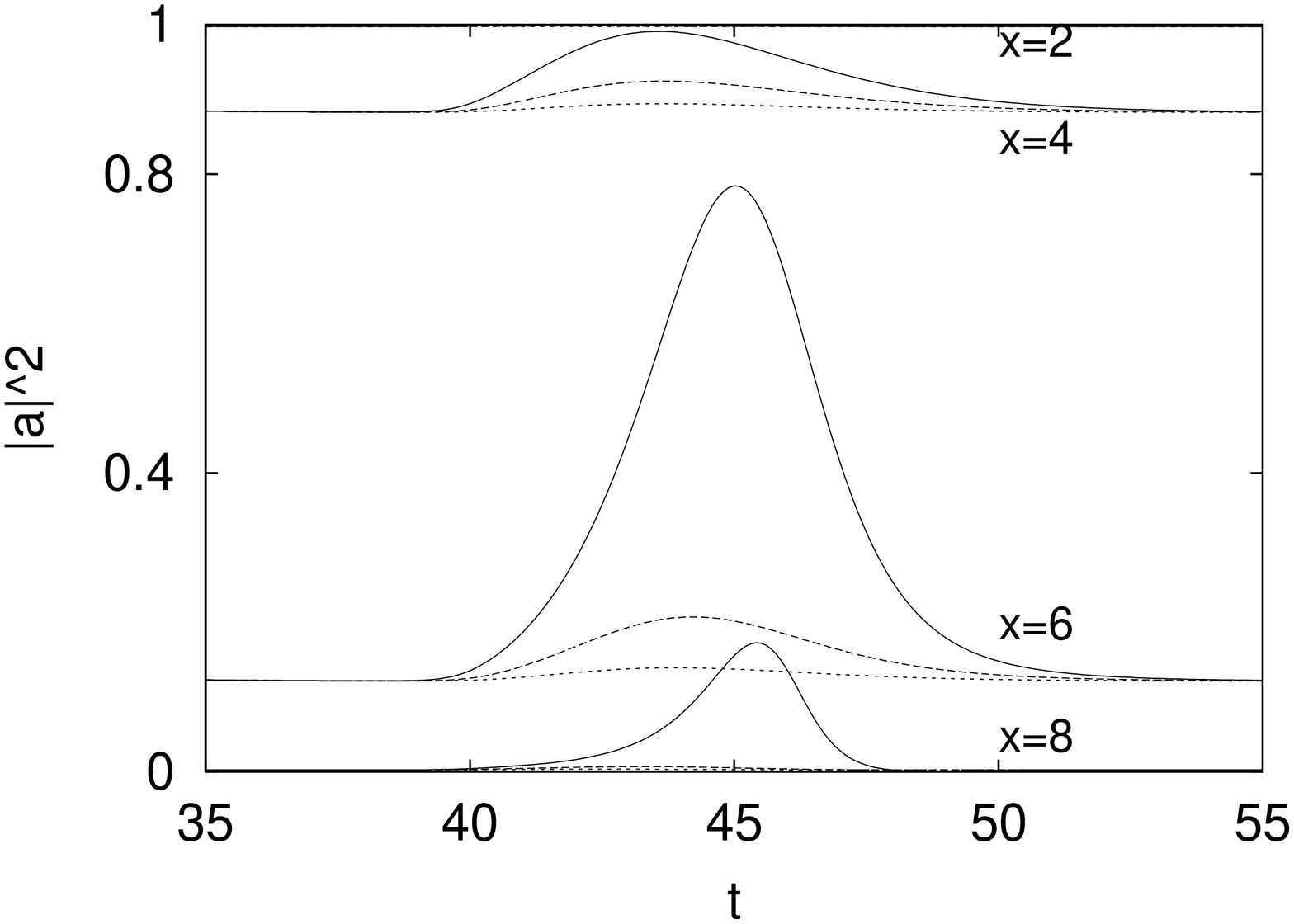,height=15 cm,width=6 cm,angle=-90}}
\caption{Influence of the amplitude of the phase change on the formation of the Raman spike.
$|a|^2 (x,t)$ for $x =2,4,6$ and $8$  as a function of $t$ for
three different phase changes of $\pi$ (full line), $\pi/2$ 
(long dash) and $\pi/4$ (short dash) occurring during a time $\delta = 1$.}
\label{f6} \end{figure}

We present here numerical solutions of the full SRS equations
for the spike which confirm the analysis done previously.
The equations (\ref{eq11},\ref{eq12}) have been integrated using the 
method indicated in section 2.
We first examine the influence of the flip time
and amplitude on the formation of the spike.

Fig. \ref{f5} shows $|a(x,t)|^2$ as a function of $t$
for $x=2,4,6$ and 8 with a Raman spike. The calculation
has been done for a flip amplitude $a_f=\pi$ and 
three values of flip time $\delta=0.1$ (full line)
1 (long dash) and 3 (short dash). As expected from the 
previous section, no manifestation of the phase flip
can be seen for 
$x<x_c={\gamma \over D^2}\approx 4$, where 
$x_c$ is the stagnation distance.
Notice how the amplitude of
the spike remains of order 1 for the fast flip $\delta=0.1$.
On the contrary for $\delta=1$ the amplitude decays as $x$
increases to 4 and 6. The decay is even stronger for 
$\delta=3$ for which the spike is barely noticeable.

The influence of the phase change on the Raman spike can be seen in
Fig. \ref{f6}. There we computed $|a (x,t)|^2$  as previously
but took $\delta=1$ fixed and took three values of the
phase jump $a_f=\pi,{\pi \over 2}$ and ${\pi \over 4}$.
The results are similar to the ones shown in Fig. \ref{f5}.
No spike can be seen for $x< x_c$ and the spike dies off for 
$\delta={\pi \over 4}$ whereas it is still weakly present
for $\delta={\pi \over 2}$.

These two figures show the importance of the ratio $a_f/\delta$. If
$a_f/\delta > \pi$ a Raman spike is created and subsists for a few stagnation
distances. On the contrary if $a_f/\delta < \pi$ the spike is weak and
short lived.

\begin{figure}[t]
\centerline{\psfig{figure=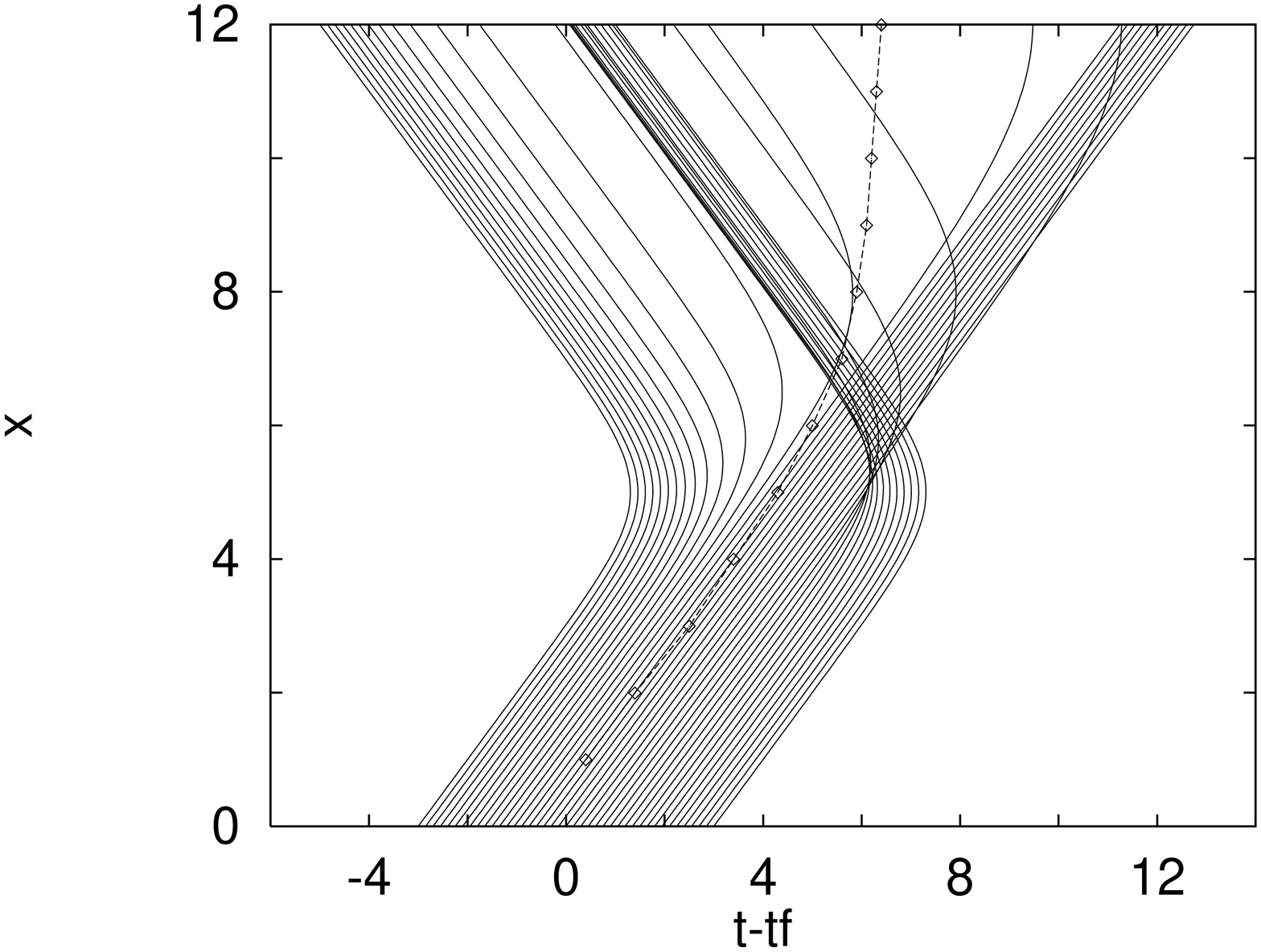,height=15 cm,width=6 cm,angle=-90}}
\caption{Characteristic curves $t(x)$ for different values of $t_0$. 
The parameters are the same as for Fig. \ref{f4}. The curve in dashed line 
indicates the position of the maximum of the spike in the 
numerical solution of the full system.}
\label{f7} \end{figure}

\begin{figure}[t]
\centerline{\psfig{figure=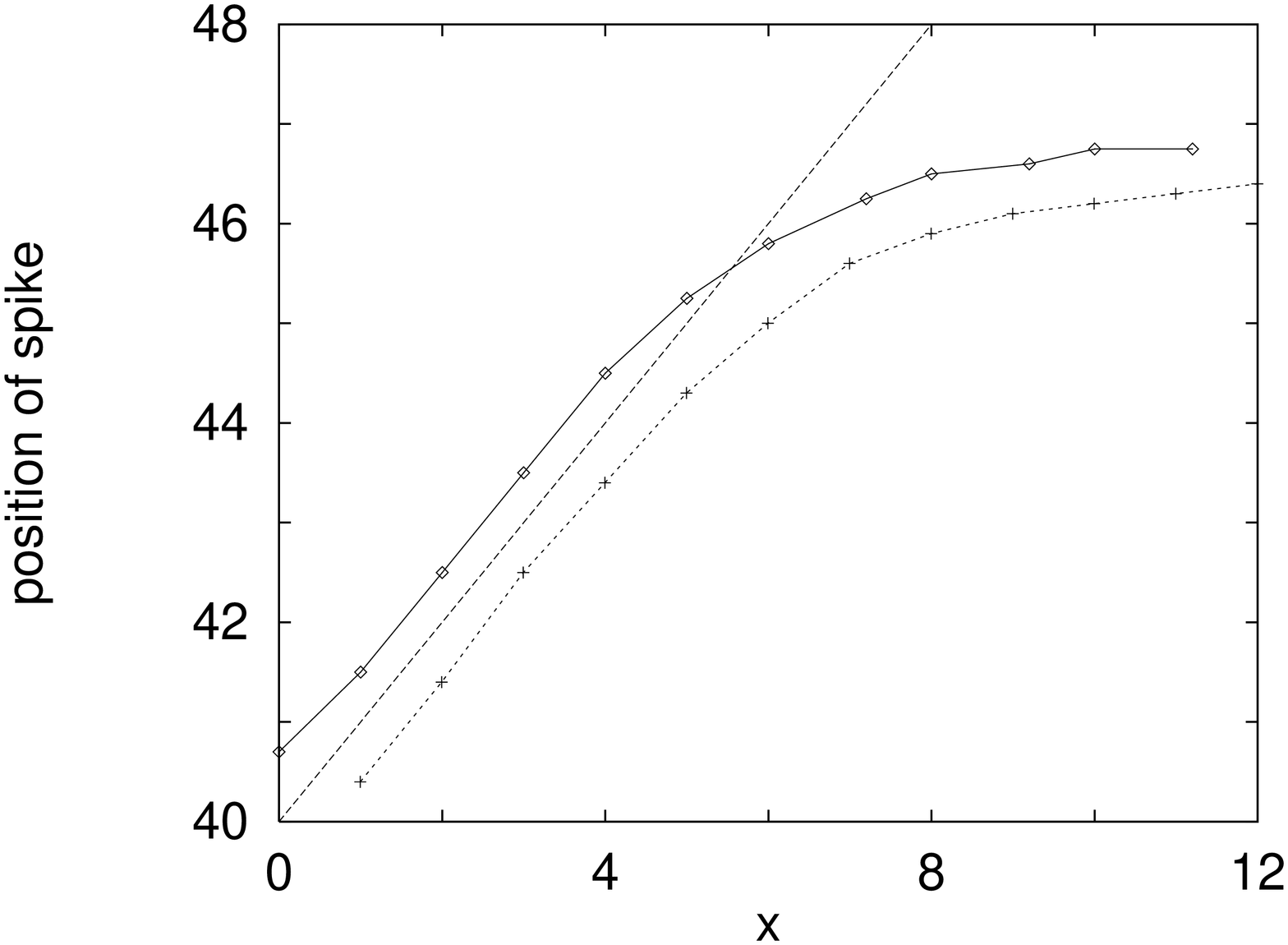,height=15 cm,width=6 cm,angle=-90}}
\caption{t position of the maximum of the Raman spike corresponding
to Fig. \ref{f4} as a function of $x$ (full line).
The $t$ position where the phase of $q$ is equal to $- \pi/2$ is 
in small dashed line. The line $t = t_f + x$ is marked in big dash.}
\label{f8} \end{figure}

\begin{figure}[t]
\centerline{\psfig{figure=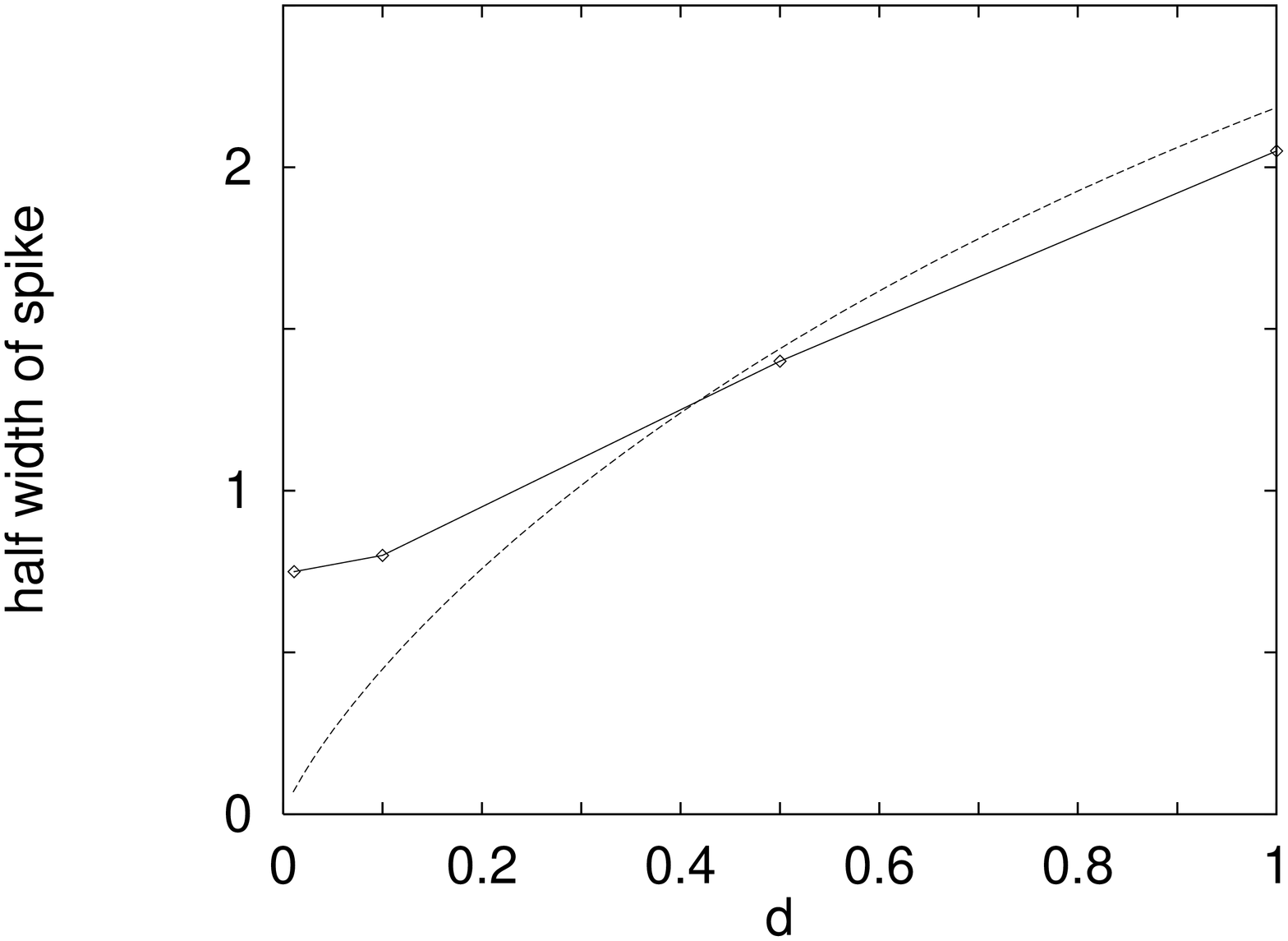,height=15 cm,width=6 cm,angle=-90}}
\caption{Half-width of Raman spike as a function of $\delta$ the 
flip time for the numerical solution of the full system (full line) for
a distance $x=2. x_c = 2 \gamma/D^2 =10 $. The analytical 
expression (\ref{eq48}) is in dashed line. The other parameters are 
the same as for Fig. \ref{f4}}
\label{f9} \end{figure}

We now compare the position and half width of the Raman spike obtained 
from the direct solution of equations (\ref{eq11},\ref{eq12}) and the estimates 
given by the characteristic equations. Fig. \ref{f7} shows the characteristic
curves $t(x)$ for different values of $t_0$ around the spike instant
$t_0=t_f$. As expected from the analysis of the previous section, one
sees the characteristic curves away from the spike region shift smoothly
from the behavior (\ref{txsmall}) to (\ref{txlarge}) as $x>x_c$. In 
the spike region
the characteristics are straight and around the spike one can see the
transition from (\ref{txsmall}) to (\ref{txlarge}) for a stagnation distance
larger than $x_c$.
Observe how the characteristics cross for $x>x_c$ indicating
a shock. In fact there are multiple shocks as shown by the
crossings seen for the curves such that $t_0 \le t_f$ in
the right top part of the graph. For $t_0 > t_f$ all
the curves seem to accumulate along a line parallel to 
(\ref{txlarge}). This set of curves shows the limitation of the
our approach which can predict the kinematics of the Raman spike
as it is created 
\be\label{kinespike}
n = -D^2 \tanh[D^2~f(t-D^2~x~\tanh(\gamma))~x -\gamma]~, \ee
but breaks down after shocks are formed.
In Fig. \ref{f7} we plot the maximum of the spike observed in the
numerical solution of (\ref{eq11},\ref{eq12}) and one can see that
this position agrees well with the one given by the accumulation
point of the characteristics for $t_0>t_f$. 
The result is clearly seen in Fig. \ref{f8} where we
plot the position of the maximum of the spike as a function of $x$.
On the same picture we give the $t^*$ value such that $q(x,t^*)=\pi/2$
and this follows the position of the spike.
In Fig. \ref{f9} we compare the width of the Raman spike with expression
(\ref{wrs}) for a range of values of $\delta$ and see that there is
a fairly good agreement.

\section{Conclusion} 

We have introduced a systematic approximation procedure to simplify
the stimulated Raman scattering equations where the small parameter is
the inverse of the damping of the field variable. Since the equations describe
complex fields, we have studied separately the influence of the 
amplitude and phase of the waves. This provided a better understanding of
the phenomenon.

Concerning the amplitude, the zero order of the approximation
yields the well-known theory for continuous waves. At the first order we
obtain
a  hyperbolic system of equations which exhibits wave breaking. At this stage
it does not give the accurate front dynamics but it could probably be
refined to do so.
The simple model we introduce allows to understand how energy transfer
occurs between the different components $a$ and $b$ of the field. In
particular we show that this transfer is monotonous and occurs
after a stagnation distance $x_c$ which we specifically compute from
the parameters of the model.

For the phase effects we provide a good description of 
the Raman spike phenomenon in the case of waves of constant 
amplitude through a simple system of hyperbolic partial differential
equations for the population difference $n= |a|^2-|b|^2$ and 
the relative phase of the fields. We show that a spike is formed 
from an initial phase jump of amplitude $a_f$
and duration $\delta$ in one of the fields if the ratio $a_f / \delta$
is large enough. From the characteristic curves of the system we obtain
the time instant and duration of the Raman spike as a function 
of the propagation distance in the medium. These estimations 
are confirmed numerically.

\section{Acknowledgements} 

A. Maimistov is grateful to the INSA de Rouen for an invited 
professorship in May 2002. J G Caputo thanks Jerome Leon for
useful discussions.


\begin{thebibliography}{9}
\bibitem{w74}  G. B. Whitham, \textit{Linear and nonlinear waves},
J. Wiley, (1974).

\bibitem{tashkent} J.\ G. Caputo, in \textit{Nonlinearity and disorder: 
theory and applications}, F.Kh. Abdullaev, M.P. Soerensen and 
O. Bang Eds. , Kluwer, (2001).

\bibitem{leon} J. Leon, Phys. Lett. A 170, 283 (1992); 
Phys.Rev. A 47, 3264 (1993).

\bibitem{lm99} J. Leon and A. Mikhailov, Phys. letters A. 1999.

\bibitem{bclp00} M. Boiti, J.G. Caputo, J. Leon et F. Pempinelli,
Inverse Problems, vol. 16, nb. 2, 303-315, (2000).

\bibitem{drulh} K. Dr\"uhl, R.G. Wenzel, J.L. Carlsten, Phys. Rev. Lett.,
{\bf51} (1983) 1171; 

\bibitem{wenzel} R.G. Wenzel, J.L. Carlsten, K. Dr\"uhl,
 J. Stat. Phys., {\bf 39} (1985) 621.

 \bibitem{yariv} A. Yariv, \textit{Quantum Electronics}
John Wiley and Sons Inc., New York (1967).

\bibitem{pantel} R.H. Pantel and H.E. Putkhof, \textit{Fundamentals of 
quantum electronics}, John Wiley and Sons Inc., New York (1969).

\bibitem{chu} F. Y. F. Chu and A. C. Scott, Phys. Rev. A 12, 2060, (1975).

\bibitem{cgl95} C.Claude, F. Ginovart and J. Leon, Phys.Rev. A 52, 767 (1995)

\bibitem{ablowitz} M.J. Ablowitz and H. Segur, \textit{Solitons and 
the Inverse Scattering Transform}, SIAM, Philadelphia, (1981).

\bibitem{newell} A.C. Newell and J.V. Moloney, \textit{Nonlinear Optics},
Addison-Wesley Publishing Company, Redwood City, (1992).

\bibitem{Agrawal95}
G.P. Agrawal, \textit{Nonlinear Fiber Optics}, Academic Press, Inc., San  Diego,
Boston, (1995).

\bibitem{Boyd} R. F. Boyd, Nonlinear optics, Academic Press, (1992).

\bibitem{BEC} G. Lenz, P.Meystre and E.W. Wright, Phys. Rev. Lett. 71, 
3271 (1993).


\end{thebibliography}
\end{document}